\documentclass[journal]{IEEEtran}

\usepackage{amsmath, amssymb, bm, cite, epsfig, psfrag, mathtools}
\usepackage{epstopdf}
\usepackage{graphicx}
\usepackage{algorithm,algpseudocode}
\usepackage{array}
\newcolumntype{L}[1]{>{\raggedright\let\newline\\\arraybackslash\hspace{0pt}}m{#1}}
\newcolumntype{C}[1]{>{\centering\let\newline\\\arraybackslash\hspace{0pt}}m{#1}}
\newcolumntype{R}[1]{>{\raggedleft\let\newline\\\arraybackslash\hspace{0pt}}m{#1}}
\usepackage{lipsum}
\usepackage{ragged2e}
\usepackage[font=small]{caption}
\usepackage{bbm}
\usepackage{multirow}
\usepackage[usenames,dvipsnames]{xcolor}
\usepackage{etoolbox}
\usepackage{pbox}
\usepackage{colortbl}
\usepackage{fancyhdr}
\usepackage{tikz}
\usetikzlibrary{calc}

\usepackage{rotating}
\usepackage{dblfloatfix}
\usepackage{supertabular}
\usepackage{longtable}
\usepackage{changepage}
\usepackage{subcaption}
\usepackage{mwe}
\usepackage{color,soul}
\usepackage{booktabs}
\usepackage{capt-of}
\usepackage{footnote}
\makesavenoteenv{tabular}
\makesavenoteenv{table}
\usepackage[hyphens]{url}
\usepackage{cite}
\usepackage{tabu}
\usepackage{fixltx2e}

\usepackage[margin=0.54in]{geometry}

\newcolumntype{P}[1]{>{\centering\hspace{0pt}}p{#1}}
\newcolumntype{M}[1]{>{\centering\hspace{0pt}}m{#1}}
\newcolumntype{L}{>{\centering\arraybackslash}m{3cm}}

\renewcommand{\arraystretch}{1.15}

\def\PL{\textrm{PL}}
\def\dB{\textrm{dB}}
\def\dBm{\textrm{dBm}}
\def\dBi{\textrm{dBi}}
\def\FSPL{\textrm{FSPL}}
\def\CI{\textrm{CI}}

\def\GHz{\textrm{GHz}}

\def\thresh{\textrm{thresh}}

\def\etal{\emph{et al. }}

\definecolor{Fgreen}{rgb}{0.13, 0.55, 0.13}

\newtoggle{conference}
\togglefalse{conference} 
\begin{document}
\bibliographystyle{IEEEtran}

\title{\LARGE Base Station Diversity Propagation Measurements at 73 GHz Millimeter-Wave for 5G  Coordinated Multipoint (CoMP) Analysis} 
\author{\IEEEauthorblockN{George R. MacCartney Jr.\textsuperscript{a}, Theodore S. Rappaport\textsuperscript{a}, and Amitava Ghosh\textsuperscript{b}}\\
\IEEEauthorblockA{\textsuperscript{a}NYU WIRELESS, \textsuperscript{b}NOKIA Bell Labs}
\vspace{-10mm}

\thanks{This material is based upon work supported by NOKIA and the NYU WIRELESS Industrial Affiliates Program, and three National Science Foundation (NSF) Research Grants: 1320472, 1302336, and 1555332. The authors thank S. Sun, Y. Xing, H. Yan, J. Koka, R. Wang, and D. Yu, for their help in conducting the measurements. Corresponding e-mail: gmac@nyu.edu}
}

\maketitle
 \begin{tikzpicture}[remember picture, overlay]
 \node at ($(current page.north) + (0,-0.25in)$) {G. R. MacCartney, Jr., T. S. Rappaport and A. Ghosh, ``Base Station Diversity Propagation Measurements at 73 GHz Millimeter-Wave};
 \node at ($(current page.north) + (0,-0.4in)$) {for 5G  Coordinated Multipoint (CoMP) Analysis," \textit{2017 IEEE Globecom Workshops (GC Wkshps)}, Singapore, Dec. 2017, pp. 1-7.};
 \end{tikzpicture}
\begin{abstract}
This paper describes wideband (1 GHz) base station diversity and coordinated multipoint (CoMP)-style large-scale measurements at 73 GHz in an urban microcell open square scenario in downtown Brooklyn, New York on the NYU campus. The measurements consisted of ten random receiver locations at pedestrian level (1.4 meters) and ten random transmitter locations at lamppost level (4.0 meters) that provided 36 individual transmitter-receiver (TX-RX) combinations. For each of the 36 radio links, extensive directional measurements were made to give insights into small-cell base station diversity at millimeter-wave (mmWave) bands. High-gain steerable horn antennas with 7$^\circ$ and 15$^\circ$ half-power beamwidths (HPBW) were used at the transmitter (TX) and receiver (RX), respectively. For each TX-RX combination, the TX antenna was scanned over a 120$^\circ$ sector and the RX antenna was scanned over the entire azimuth plane at the strongest RX elevation plane and two other elevation planes on both sides of the strongest elevation angle, separated by the 15$^\circ$ HPBW. Directional and omnidirectional path loss models were derived and match well with the literature. Signal reception probabilities derived from the measurements for one to five base stations that served a single RX location show significant coverage improvement over all potential beamformed RX antenna pointing angles. CDFs for nearest neighbor and Best-\emph{N} omnidirectional path loss and cell outage probabilities for directional antennas provide insights into coverage and interference for future mmWave small-cells that will exploit macro-diversity and CoMP.  
\end{abstract}

\iftoggle{conference}{}{
\begin{IEEEkeywords}
MmWave, channel sounder, 73 GHz, diversity, macro-diversity, CoMP, outage, path loss, beamforming. 
\end{IEEEkeywords}}

\section{Introduction}\label{sec:intro}
Millimeter-wave (mmWave) bands will play an important role in fifth-generation (5G) wireless communications for mobile access and backhaul~\cite{Rap13a,Boccardi14a}. 5G trials are on-going for fixed-wireless and mobile scenarios in mmWave bands, with customer trials expanding~\cite{Fortune17a}. Numerous propagation measurements were performed to show that mmWave bands are viable for future 5G wireless and to assist in the creation of channel models~\cite{Rap15b,A5GCM15,3GPP.38.901,mmMAGIC16a}. However, a few key aspects have yet to be extensively examined at mmWave bands such as spatial consistency, dynamic blocking and rapid fading, and base station diversity~\cite{Mac16a,Samuylov16a,Rap17a,Rap17b}. This paper presents initial results from a coordinated multipoint (CoMP)-style propagation measurement campaign conducted at NYU during the summer of 2016, with the goal of understanding mmWave base station diversity characteristics, also known as macro-diversity. 

Base station and/or antenna diversity in small cells is well understood to improve signal-to-interference-plus-noise-ratio (SINR) for increased capacity and to reduce outage~\cite{Foschini06a}. CoMP techniques were standardized by 3GPP and are employed in Long Term Evolution-Advanced (LTE-A) to make use of spatially separated base stations to jointly communicate with a mobile user through coordination~\cite{3GPP.36.819,Lee12a}. CoMP is typically referred to as network or distributed multiple-input multiple-output (MIMO) and reduces intercell interference, increases network capacity, and improves coverage for cell-edge users~\cite{Foschini06a}. Base station diversity and CoMP are attractive technologies for mmWave systems in order to reduce outages caused by dynamic channel conditions and human blocking~\cite{Mac16a,Rap17b}. 

MmWave systems will use high-gain and fast-switching electrically steerable antennas to overcome the additional free space path loss (FSPL) in the first meter of propagation compared to sub-6 GHz bands~\cite{Sun14b}. An implication of this architecture is the influence of dynamic objects in both line-of-sight (LOS) and non-LOS (NLOS) scenarios that can cause deep fading in beamformed links between a base station and mobile. Traditional cellular systems use quasi-omnidirectional or sectored antennas at the base station and nearly omnidirectional antennas at the mobile (also called user equipment or UE) that result in broad angles of departure and arrival. 

There are very few large-scale propagation studies for CoMP / base station diversity in dense urban environments at mmWave frequencies. Measurements were conducted by Le \etal at 25 GHz over 20 MHz of bandwidth and at 38 GHz over 200 MHz of bandwidth for links less than 1 kilometer (km) to see how diversity could overcome adverse effects of rain at centimeter-wave (cmWave) and mmWave bands~\cite{Le14a}. Results in~\cite{Le14a} with 29 dBi and 32 dBi high-gain antennas showed that as the spatial separation between base stations increased, site diversity gain increased. A second observation corroborated earlier measurement results and theory where diversity gain increased as angle separation between the mobile and two base stations increased~\cite{Le14a}.

A 5G multipoint field trial was conducted at 15 GHz in the parking lot of NTT Docomo in Japan with a 5G prototype system in a 100 m x 70 m cell and a 20 m x 20 m small-cell to investigate throughput gains using multipoint coordination between two transmission points (TPs) each with 90$^\circ$ sectored-beam antennas, as a UE with multiple omnidirectional antennas traveled along a route~\cite{Kurita15a}. Downlink CoMP with joint transmission (JT) and dynamic point selection (DPS) revealed a 70\% throughput gain in the small-cell study and a 30\% throughput improvement in the larger study area, compared to a single serving TP. 

Ericsson conducted a study in Stockholm, Sweden to test distributed MIMO and CoMP potentials with 4x4 MIMO at 15 GHz over 200 MHz of bandwidth and with two base station TPs that each consisted of an antenna array with 15 dBi and 90$^\circ$ HBPW antennas while the UE had an array of omnidirectional antennas~\cite{Halvarsson16a}. Spatial multiplexing gains were achieved in LOS where throughput improved from 1.1 Gbps (5.7 bps/Hz) to 2.5 Gbps (12.6 bps/Hz). For large angular separation between the two TPs and the UE (appears to be 180$^\circ$), rank-4 transmission was consistent with a maximum achievable spectral efficiency of 13 bps/Hz~\cite{Halvarsson16a}.

In order to obtain an understanding of the mmWave channel for real-world base station diversity and CoMP applications, we conducted a large-scale measurement campaign using directional high-gain and narrowbeam antennas which is described in Section~\ref{sec:meas}. Preliminary results and models are presented in Section~\ref{sec:prelim}, and Section~\ref{sec:conc} provides conclusions.

\section{Measurement Campaign and Hardware}\label{sec:meas}
\subsection{Measurement Environment}
\begin{figure}
	\centering
	\includegraphics[width=0.48\textwidth]{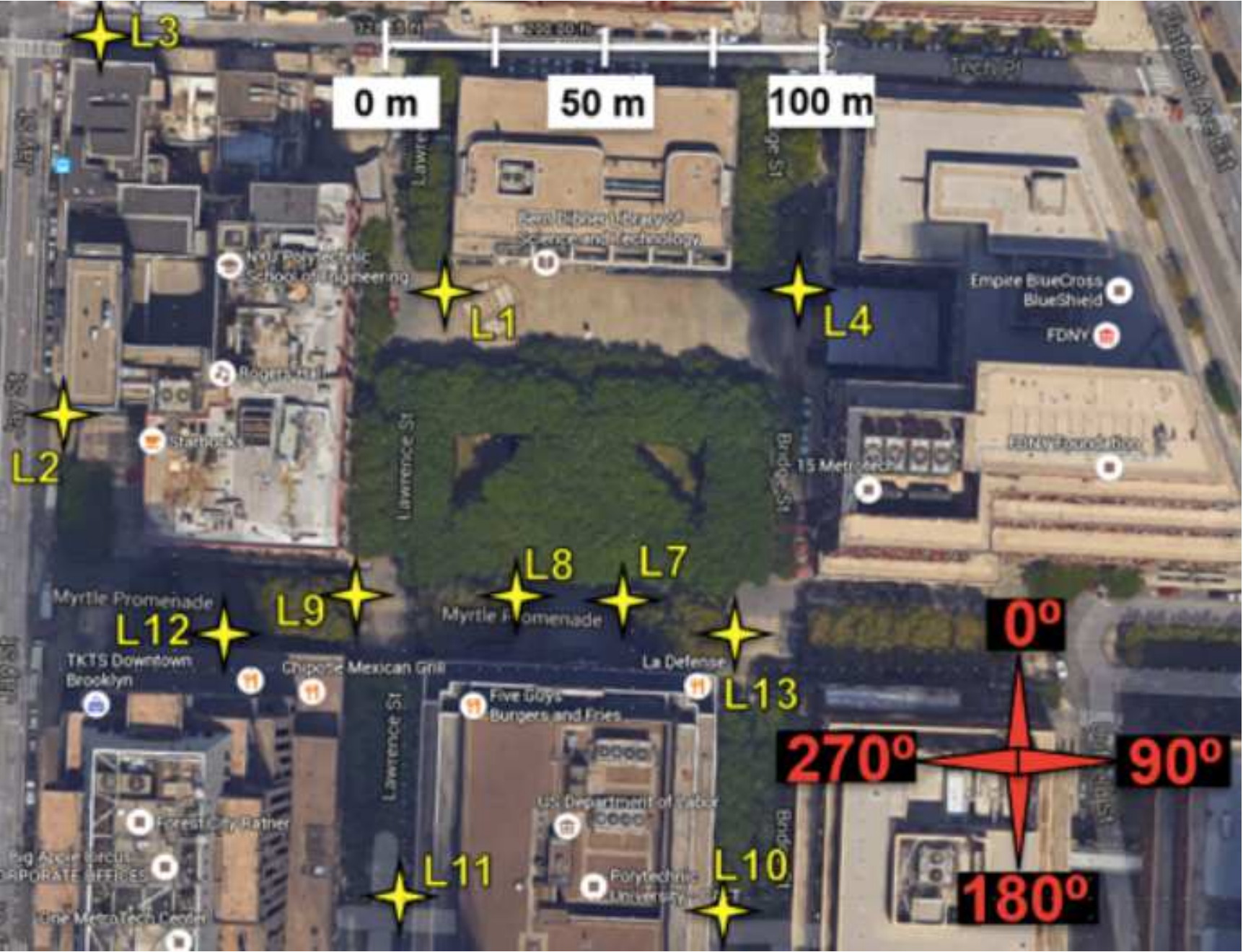}
	\caption{TX and RX locations on the NYU Engineering campus. The measurement area is approximately 200 m by 200 m and Table~\ref{tbl:TXRXlocs} indicates the TX-RX combinations measured.}\label{fig:Map}
\end{figure}

Propagation measurements at 73.5 GHz were conducted in summer 2016 on the NYU Engineering campus, which is built around an orchard of cherry trees that is surrounded by an open square (O.S.) in downtown Brooklyn, New York. The campus was a typical downtown environment with lampposts, street signs, tall buildings, walkways, foliage, etc., and the study area spanned $\sim$200 m by 200 m, representing an urban microcell (UMi) scenario. The campus is surrounded by buildings of four to fifteen stories on all four sides, with urban canyon streets at the intersection of each corner. Overall, 11 locations were chosen for both TX and RX locations, representing typical locations for base stations or heavy-user concentrations, based on access to power and are displayed as yellow stars in Fig.~\ref{fig:Map}. Of the 11 locations, on any given day, one of the locations was chosen as the TX location, and some of the other 10 locations would be used for measuring RX responses. Each of the 11 locations was systematically chosen as a TX location when the other locations were used as RX locations throughout the measurement campaign. Not all RX locations were measured for a given TX, but the goal was to measure at least 3 RX's for each TX and to transmit from at least 3 TX's to each RX, for a diverse range of nearest neighbor distances and environments for each measurement subset. Over 150 Gigabytes (GBs) of data were collected for directional scanning antenna measurements as described below. In total, 36 TX-RX radio links were tested, where signals were received from 3 or more TX locations at 9 RX locations, and where 8 TX locations were used to transmit to 3 or more RX locations. Measurements revealed values and variations in received power, outage, best beam angles, and temporal delay spread among various pointing angles at each RX, for the case of multiple serving base station locations and pointing directions in both LOS and NLOS.

A total of 11 LOS and 25 NLOS TX-RX links were measured using 14.9 dBm TX power (into the antenna) and 1 GHz of RF null-to-null bandwidth, with 3D transmitter-receiver (T-R) separation distances that ranged from 21 m to 140 m in LOS and 59 m to 170 m in NLOS. The 3D T-R separation distances were measured as the 3D Euclidean distance between a TX and RX, where TX heights were 4.0 m above ground level (AGL) and RX heights were 1.4 m AGL. The heights were chosen to emulate a small-cell scenario where access points (APs) are on lampposts and very tall street signs and to emulate a mobile user interacting with their device in an Internet browsing mode. Throughout this paper, the terms \textit{base station (BS)} and \textit{access point (AP)} are used interchangeably. 

Table~\ref{tbl:TXRXlocs} indicates the TX-RX location combinations and their corresponding T-R separation distance ranges. Table~\ref{tbl:NNdistances} provides the mean, median, standard deviation, and range of distances ($R$) of the nearest serving base stations for the RX locations. The green patches in the center of the open square in Fig.~\ref{fig:Map} consist of moderate to full foliage, with the canopy approximately 6-7 meters AGL. The 4 meter TX heights were below the tree canopy and thus only thin branches and tree trunks contributed to light obstructions in propagation paths. A TX-RX combination was considered to be LOS if a straight line (in 3D space) drawn between the TX antenna and RX antenna resulted in a clear optical path. If a clear optical path was obstructed by either a large tree trunk or buildings, then the TX-RX combination was considered NLOS. In cases where a TX-RX combination had light branches along the straight line drawn between the TX and RX antennas, the setting was specified as LOS.

\begin{table}
\caption{List of the TX locations used to transmit to each RX location.}
\label{tbl:TXRXlocs}
	\begin{center}
	\scalebox{0.98}{\footnotesize
		\begin{tabular}{|c|c|c|}
			\hline
			RX Location &	Serving TX Location	&	T-R Dist. Range (m) \\ \specialrule{1.5pt}{0pt}{0pt}
			L1	&	L3,L4,L7,L11,L13	&	80 $\leq d \leq 140$ \\ \hline
			L2	&	L3,L9,L12			&	61 $\leq d \leq 78$ \\ \hline
			L3	&	L2					& 	77 \\ \hline
			L4	&	L1,L3,L7,L10,L13	&	80 $\leq d \leq 170$ \\ \hline
			L7	&	L1,L2,L4,L10		&	72 $\leq d \leq 133$ \\ \hline
			L8	&	L1,L7,L9			&	21 $\leq d \leq 78$ \\ \hline
			L9	&	L1,L2,L4,L11		&	63 $\leq d \leq 123$ \\ \hline
			L10	&	L4,L7,L13			&	59 $\leq d \leq 140$ \\ \hline
			L12	&	L1,L2,L4,L7,L11		&	61 $\leq d \leq 149$ \\ \hline
			L13	&	L1,L4,L10			&	59 $\leq d \leq 107$ \\ \hline
		\end{tabular}}
	\end{center}
\end{table}
\begin{table}
	\centering
	\caption{Distance ($R$) statistics for the 9 UMi RX locations and their nearest neighbor TX locations, rounded to the nearest integer.} \label{tbl:NNdistances}
	\scalebox{0.81}{
		{\renewcommand{\arraystretch}{1.3}
			\begin{tabular}{|c||c|c|c|}  \hline
				\textbf{Nearest Neighbor}		& \textbf{1}	& \textbf{2}	& \textbf{3}\\	\specialrule{1.5pt}{0pt}{0pt}
				Mean [m]: $\bar{R}$ / Median [m]: $\tilde{R}$& 63/63		& 75/78			& 93/87			\\ \hline
				STD [m]: $\sigma_{R}$			& 19			& 14			& 20			\\ \hline
				Range [m]: $R\in[\min,\max]$	& [21,80]		& [41,87]		& [78,140]		\\ \hline
			\end{tabular}}}
		\end{table}

\subsection{Measurement Hardware}
The measurement campaign was conducted using an absolute-timing ultra-wideband sliding correlator channel sounder that transmitted a 500 Mega-chips-per-second (Mcps) pseudorandom noise (PN) sequence with a 1 GHz RF null-to-null bandwidth at a center frequency of 73.5 GHz~\cite{Mac17a,Mac17c}. The system employed narrowbeam rotatable directional horn antennas with a 7$^\circ$ azimuth/elevation (Az./El.) half-power beamwidth (HPBW) and 27 dBi of gain at the TX, and a 15$^\circ$ Az./El. HPBW antenna with 20 dBi of gain at the RX. Narrowbeam and directional antennas were used at both the TX and RX for spatial filtering and to emulate future mmWave systems that will consist of adaptable directional beams from a base station to somewhat broader beamwidths that are steerable at a mobile handset. The TX and RX antennas were mechanically rotated by programming a FLIR D100 pan/tilt gimbal in LabVIEW. The channel sounder was able to record absolute and relative time-delay power delay profiles (PDPs) per the synchronization procedure outlined in~\cite{Mac17a}, with 2 ns multipath component (MPC) time resolution. The synchronization procedure allowed for synthesizing accurate omnidirectional and sectored PDPs. Table~\ref{tbl:sysSpec} provides specifications for the ultra-wideband sliding correlator channel sounder (See \cite{Mac17a} for additional details). 

\begin{table}
	\centering
	\caption{Ultra-wideband sliding correlator channel sounding system specifications used for the 73 GHz CoMP/AP diversity campaign~\cite{Mac17a}.}
	\label{tbl:sysSpec}
	\begin{center}
		\scalebox{0.90}{
			\linespread{0.99}
			\fontsize{9.0}{9.0}\selectfont
			\begin{tabular}{|c|c|}
				\hline 
				\textbf{Carrier Frequency} 				& \textbf{73.5 GHz} 		\tabularnewline \specialrule{1.5pt}{0pt}{0pt}
				\textbf{Probing Signal} 				& 11\textsuperscript{th} order PRBS (length=2047)	\tabularnewline \hline
				\textbf{TX PN Code Chip Rate} 			& 500 Mcps 					\tabularnewline \hline
				\textbf{TX PN Code Chip Width} 			& 2.0 ns 					\tabularnewline \hline
				\textbf{RX PN Code Chip Rate}			& 499.9375 Mcps 			\tabularnewline \hline
				\textbf{Slide Factor} 					& $8\,000$ 					\tabularnewline \hline
				\textbf{RF Center Frequency}			& 73.5 GHz					\tabularnewline \hline
				\textbf{RF Bandwidth (Null-to-Null)} 	& 1 GHz						\tabularnewline \hline
				\textbf{TX Antenna Gain} 				& 27 dBi 					\tabularnewline \hline
				\textbf{TX Az./El HPBW} 				& 7$^{\circ}$/7$^{\circ}$	\tabularnewline \hline
				\textbf{RX Antenna Gain} 				& 20 dBi 					\tabularnewline \hline
				\textbf{RX Az./El. HPBW} 				& 15$^{\circ}$/15$^{\circ}$	\tabularnewline \hline
				\textbf{Max. TX Power / Max. EIRP} 		& 14.9 dBm / 41.9 dBm \tabularnewline \hline
				\textbf{TX/RX Antenna Height} 			& 4.0 m / 1.4 m 			\tabularnewline \hline  
				\textbf{Max. Measurable Path Loss}		& 175 dB w/ antennas		\tabularnewline \hline
				\textbf{Multipath Time Resolution} 		& 2 ns 						\tabularnewline \hline 
				\textbf{TX Polarization} 				& Vertical					\tabularnewline \hline
				\textbf{RX Polarization} 				& Vertical / Horizontal		\tabularnewline \hline
		\end{tabular}}
	\end{center}
\end{table}

\subsection{Measurement Methodology and Beam Sweeping Procedure}
To understand the directional nature of the mmWave channel and to obtain a large amount spatial information at the TX and RX for AP diversity analysis, the transmitted signal for each TX-RX combination was radiated with a 7$^\circ$ HPBW antenna and was rotated in sequential increments of 8$^\circ$ to span at least a 120$^\circ$ sector (15 pointing angles), representative of a typical panel base station. For six of the measured TX-RX links with rich reflections and scattering observed beyond the 120$^\circ$ sector, a larger TX sector of 136$^\circ$ was measured. For each TX antenna pointing angle, the RX antenna with a 15$^\circ$ HPBW antenna was scanned in 15$^\circ$ increments over the complete azimuth plane and this was repeated for three elevation planes. Measurements were recorded for the RX elevation angle with the strongest received power and $\pm$ 15$^\circ$ from the strongest elevation angle which was typically 2$^\circ$ to 5$^\circ$ above the horizon. A PDP was recorded with 20 averages for each unique antenna pointing angle combination between the TX and RX where a signal was detectable. If the signal was not detectable, a PDP was not recorded.

For three elevation planes and 15 TX pointing angles, this amounted to 45 azimuthal measurement scans for co-polarized vertical-to-vertical (V-V) antennas. The same procedure was repeated for cross-polarized vertical-to-horizontal (V-H) antennas, resulting in a maximum of 90 azimuthal scans for a single TX-RX combination. As described earlier, there were six instances where 2 additional TX pointing angles were measured, resulting in 102 scans. Each of the azimuthal scans consisted of 24 recorded PDPs, for a maximum of 2,160 recorded PDPs per TX-RX location combination (2,448 for 102 azimuth scans), although, 700 PDPs were recorded on average for each TX-RX location combination since angles where a signal was not detectable were not recorded. A total of 25,614 PDPs were recorded. We note that the start (1st) and end (15th) angles of the TX sector were manually determined during the measurements for each TX-RX link so as to measure the most dominant angles of departure that resulted in the strongest received power at the RX. The center angle (8th) of the sector typically resulted in the maximum received power at the RX.

\section{Preliminary Results}\label{sec:prelim}
\subsection{Path Loss Models}
UMi O.S. directional and omnidirectional path loss models were by-products of the measurements. Path loss models describe large-scale propagation loss for link budget and interference analysis. Friis' free space transmission formula describes the received power at a distance $d$ in meters as calculated by~\cite{Friis46a}:
\begin{equation}\label{eq:FriisLin}
\footnotesize
P_r(d) = P_t G_t G_r \left(\frac{\lambda}{4\pi d}\right)^2
\end{equation}
where $P_r$ is the received power in milliwatts (mW), $P_t$ is the transmit power in mW, $G_t$ and $G_r$ are the linear gains of the TX and RX antennas, respectively, $\lambda$ is the wavelength in meters, and $d$ is the 3D Euclidean distance in meters between the TX and RX. Friis' transmission formula in~\eqref{eq:FriisLin} is often re-written in log-scale as~\cite{Friis46a,Rap15b}:
\begin{equation}\label{eq:FriisdB}
\footnotesize
P_r(d)[\dBm] = P_t[\dBm]+G_t[\dBi]+G_r[\dBi]+20\log_{10}\left(\frac{\lambda}{4\pi d} \right)
\end{equation}
The square exponent in~\eqref{eq:FriisLin} and the ``$20$" before the log-term in~\eqref{eq:FriisdB} indicate that radio waves decay by 20 dB per decade of distance in free space. 

An interesting note about~\eqref{eq:FriisLin} and Friis' transmission formula is the relationship between antenna gain, the size of the antenna aperture, and the carrier wavelength. In general, antenna gain is a function of the effective aperture area $A_e$ and wavelength $\lambda$: $G = \frac{A_e 4\pi}{\lambda^2}$.
If one increases the frequency and keeps the same physical size of the antenna aperture at the TX and RX, then received power in free space is greater at higher frequencies than at lower frequencies. This received power gain in free space can be expressed as a function of the two frequencies:
\begin{equation}\label{eq:gainIncrease}
\footnotesize
G_{\mathrm{increase}} = \left(\frac{f_2}{f_1}\right)^2
\end{equation}
where $f_1$ is the lower frequency and $f_2$ is the higher frequency, with identical size TX and RX antenna apertures at both frequencies~\cite{Rap15a}.

Regardless of the use of directional or omnidirectional antennas, received power in the real-world (non-free space channels) can be calculated as a function of the measured path loss when removing the antenna gains from the path loss model~\cite{Rap15a}:
\begin{equation}\label{eq:PLant}
\footnotesize
P_r(d)[\dBm] = P_t[\dBm]+G_t+G_r-\PL(d)[\dB]
\end{equation}
where $\PL(d)[\dB]$ is the observed path loss at a 3D distance $d$.

The close-in free space reference distance (CI) path loss model is commonly used in the literature and standard bodies and is written as~\cite{Rap15b,Sun16b}:
\begin{equation}\label{eq:CI1}
\footnotesize
\PL(f_c,d)[\dB] = \FSPL(f_c,d_0)+10n\log_{10}\left(\frac{d}{d_0}\right)+\chi_\sigma\text{ for }d\geq d_0
\end{equation}
where $f_c$ is the carrier frequency in Hz, $d_0$ is a reference distance (1 m), $n$ is the path loss exponent (PLE), and $\FSPL(f_c,d_0)$ is the FSPL at 1 m at $f_c$ given by: $20\log_{10}\left(\frac{4\pi}{\lambda}\right)=32.4+20\log_{10}\left(\frac{f_c}{\mathrm{1 GHz}}\right)$. The zero-mean Gaussian random variable $\chi_\sigma$ with standard deviation $\sigma$ (in dB) represents the shadow fading.  A common way of re-writing~\eqref{eq:CI1} is in the 3GPP-style format with $d_0$ set to 1 m~\cite{Mac17d,Sun16b}:
\begin{equation}\label{eq:CI2}
\footnotesize
\PL^{\CI}(f_c,d)[\dB]= 32.4+10n\log_{10}(d)+20\log_{10}(f_c)+\chi_\sigma
\end{equation}
where $d$ is the 3D Euclidean distance between the TX and RX, $f_c$ is the carrier frequency in GHz, $n$ is the PLE without antenna gains considered in path loss computation~\cite{Sulyman14a,Rap15b,Mac15b}, and 32.4 is FSPL at 1 GHz at 1 m. The use of 1 m as a reference distance allows path loss to be tied to a true physical anchor point that represents the free space transmit power away from the TX antenna and at a close-in reference distance $d_0$ where no obstructions or blockages are likely to exist. 

\subsubsection{UMi Open Square Directional Path Loss}
The directional path loss models are based on the AP diversity measurements performed across numerous angles of departure (AODs) and angles of arrival (AOAs) as described in Section~\ref{sec:meas}. The path loss data was calculated by integrating the power under each PDP and subtracting that from the transmit power and removing the TX and RX antenna gains as indicated in~\eqref{eq:PLant}, for each of the arbitrary antenna pointing angles between the TX and RX. FSPL at the close-in reference distance of $d_0$ = 1 m at 73.5 GHz is 69.8 dB. Table IV in~\cite{Rap15b} describes the environmental designation terminology for the LOS, NLOS, and NLOS-best \textit{directional} path loss models~\cite{Rap15b} given subsequently.

\begin{table}
	\centering
	\caption{73 GHz directional CI V-V path loss model~\eqref{eq:CI2} parameters for the UMi O.S. scenario with TX heights of 4.0 m and RX heights of 1.4 m. TX antennas had 27 dBi of gain with 7$^\circ$ Az./El. HPBW and RX antennas had 20 dBi of gain with 15$^\circ$ Az./El HPBW.} \label{tbl:Dir_PL_VV_CI}
	\fontsize{8}{8}\selectfont
	\scalebox{0.97}{
		\begin{tabu}{|c|c|[1.6pt]c|c|[1.6pt]c|c|[1.6pt]c|c|}  \hline
			\multicolumn{8}{|c|}{\textbf{73 GHz Directional CI Path Loss Models for $\bm{d_0 = 1}$ m}} \\ \specialrule{1.5pt}{0pt}{0pt}
			\multirow{2}{*}{Freq.}	& \multirow{2}{*}{Pol.}& \multicolumn{2}{c|[1.6pt]}{LOS} & \multicolumn{2}{c|[1.6pt]}{NLOS} & \multicolumn{2}{c|}{NLOS-best} \\ \cline{3-8}
			& & PLE & $\sigma$ [dB] & PLE & $\sigma$ [dB] & PLE & $\sigma$ [dB]\tabularnewline \specialrule{1.5pt}{0pt}{0pt}
			73 GHz	& V-V & 2.0		& 1.9	& 4.6	& 11.4	& 2.9	& 11.0	\\ \specialrule{1.5pt}{0pt}{0pt}
	\end{tabu}}
\end{table}
Table~\ref{tbl:Dir_PL_VV_CI} provides the CI V-V path loss model parameters for a 1 meter reference distance derived from the directional measurements. The LOS CI path loss model has a PLE of $n$ = 2.0 which perfectly matches theoretical FSPL by Friis' free space transmission formula~\cite{Friis46a}. The NLOS CI PLE of $n$ = 4.6 indicates that 73 GHz radio frequencies attenuate by 46 dB per decade of distance beyond 1 m in the UMi O.S. scenario, which is nearly identical to the PLE of 4.7 from UMi measurements in Manhattan~\cite{Rap15b}. A surprising result here is the PLE of $n$ = 2.9 for the best TX-RX pointing angle combinations in NLOS (\emph{NLOS-best}). This result shows that if the TX and RX antennas can beamform and optimally align, that path loss can be reduced in NLOS by 18 dB per decade of distance beyond the first meter of propagation, compared to arbitrary antenna pointing angles. This observation is larger than the 7 dB and 10 dB per decade improvement for UMi in Manhattan at 28 GHz and 73 GHz, respectively~\cite{Rap15b}. This improvement in link margin for the best TX-RX pointing angle combinations may be attributed to the much more extensive beam sweeping conducted in this campaign or might be due to the fact that the measured environment has buildings on all four sides of the square that support rich reflections and scatterers.

\subsubsection{UMi Open Square Omnidirectional Path Loss}
While directional path loss models are useful for simulating and designing systems that will use narrowbeam antennas and beamforming techniques, omnidirectional path loss models are historically used by standards bodies since most legacy wireless systems use quasi-omnidirectional or sectored antennas at the base station and relatively omnidirectional antennas at the mobile side. Typically, arbitrary antenna patterns and MIMO processing are simulated with omnidirectional path loss models, but accurate representation of the angles and time delays of the channel must be used~\cite{Sun17b}. The method for synthesizing omnidirectional path loss values from directional path loss data described in~\cite{Rap15b,Mac15b,Sun15a} was performed with the directional path loss data. The LOS and NLOS terminology for the omnidirectional path loss models is provided in Table~\ref{tbl:omniPLdescrip}.

\begin{figure}
	\centering
	\includegraphics[width=0.44\textwidth]{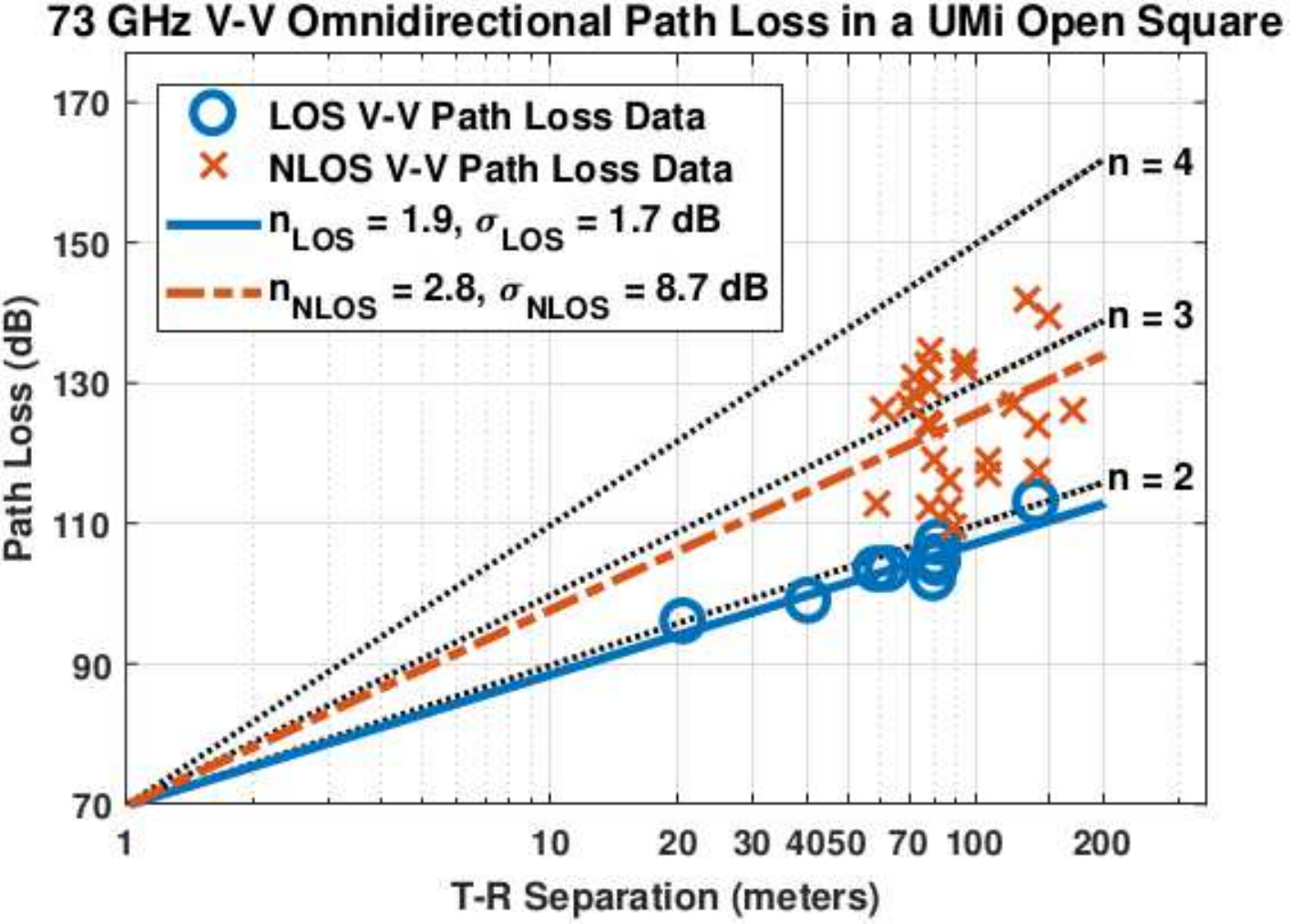}
	\caption{73 GHz omnidirectional CI V-V path loss for the UMi O.S. scenario in downtown Brooklyn. Blue circles represent LOS path loss data and red crosses indicate NLOS path loss data}\label{fig:OmniPL}
\end{figure}

The LOS and NLOS CI V-V omnidirectional path loss models~\eqref{eq:CI2} and path loss data are plotted in Fig.~\ref{fig:OmniPL}, with CI path loss model parameters given in Table~\ref{tbl:Omni_PL_VV_CI}. The omnidirectional CI model PLE is $n$ = 1.9 and indicates a multipath rich LOS environment. Since Table~\ref{tbl:Omni_PL_VV_CI} indicated that the directional LOS PLE was $n$ = 2.0, an omnidirectional PLE of $n$ = 1.9 indicates that the LOS UMi O.S. environment results in 1 dB less attenuation per decade of distance when using omnidirectional antennas compared to directional antennas. However, it is important to note that omnidirectional antennas have less gain than directional antennas, meaning the system range and link budget will be smaller when using omnidirectional antennas for a given TX power level into the antenna, despite the smaller PLE~\cite{Rap13b}. Thus, it can be inferred that the direct LOS path contains most but not all of the propagating energy from the TX to the RX such that multipath reflections, scatterers, and diffuse components also contribute to the total received power at the RX. The omnidirectional LOS PLE of $n$ = 1.9 is comparable to the LOS PLE of $n$ = 1.85 reported for the UMi O.S. scenario in~\cite{A5GCM15}.

\begin{table}
	\begin{center}
		\centering
		\renewcommand{\arraystretch}{1.3}
		\caption{Omnidirectional path loss model environment terminology.} \label{tbl:omniPLdescrip}
		\fontsize{8.5}{8.5}\selectfont
		\begin{tabular}{ | c || p{6.2cm} |}
			\hline
			\textbf{Setting} & \textbf{Description}\\ \hline \cline{1-2}
			\textbf{LOS} & Path loss when there is a clear optical path between the TX and RX site. \\ \hline
			\textbf{NLOS} & Path loss when the TX and RX sites are separated by obstructions and there is no clear direct/optical path between the TX and RX sites. \\ \hline
		\end{tabular}
	\end{center}
\end{table}
\begin{table}
	\centering
	\caption{73 GHz omnidirectional CI V-V path loss model~\eqref{eq:CI2} parameters with $d_0 = 1$ for the UMi O.S. scenario in downtown Brooklyn with TX heights of 4.0 m and RX heights of 1.4 m.} \label{tbl:Omni_PL_VV_CI}
	\scalebox{1.00}{
	\begin{tabu}{|c|c|[1.6pt]c|c|[1.6pt]c|c|}  \hline
		\multicolumn{6}{|c|}{\textbf{73 GHz Omnidirectional CI Path Loss Models for $\bm{d_0 = 1}$ m}} \\ \specialrule{1.5pt}{0pt}{0pt}
		\multirow{2}{*}{Freq.}	& \multirow{2}{*}{Pol.}& \multicolumn{2}{c|[1.6pt]}{LOS} & \multicolumn{2}{c|}{NLOS} \\ \cline{3-6}
		& & PLE & $\sigma$ [dB] & PLE & $\sigma$ [dB] \\ \specialrule{1.5pt}{0pt}{0pt}
		73 GHz	& V-V & 1.9		& 1.7	& 2.8	& 8.7	\\ \specialrule{1.5pt}{0pt}{0pt}
	\end{tabu}}
\end{table}

The NLOS PLE for the UMi O.S. at 73 GHz was determined to be $n$ = 2.8 and might indicate more reflections and scattering compared to the UMi Street Canyon and UMa scenarios that reported PLEs of $n$ = 3.0 or greater~\cite{A5GCM15}. The relatively low PLE of 2.8 in NLOS in the UMi O.S. might be attributed to light foliage and large cross-sectional reflectors (buildings) in the environment that result in low attenuation and strong multipath components, respectively. The UMi O.S. NLOS CI path loss model in~\cite{A5GCM15} reported a PLE of 2.83 which is nearly identical to the PLE of 2.8 found here.

\subsection{Directional Outage Probabilities}
For each of the TX locations indicated in Table~\ref{tbl:TXRXlocs}, the RX locations measured for each TX may be considered as random drops in a realistic small-cell scenario with multiple base stations over an approximate 200 m by 200 m area. We first note that received power was measured for at least one arbitrary TX-RX antenna pointing angle combination for each of the 36 TX-RX links, resulting in a 0\% outage for the entire campaign. This means that there was at least one TX and RX pointing angle for every TX-RX combination that enabled signal reception above a 5 dB SNR threshold, which is remarkable given that the TX power was only 14.9 dBm into the TX antenna. Also note that an outage means that received power was not measurable above the 5 dB SNR threshold of our system when recording PDPs, for which the maximum measurable path loss was 175 dB with the TX/RX antennas used (see Table~\ref{tbl:sysSpec}). Since each RX was scanned in 15$^\circ$ increments over the $360^\circ$ azimuth plane (24 pointing angles) and at three separate elevation planes, 72 overall RX beamformed angles were tested. In order to learn the outage for the beamformed angles at each RX and over all of the TX antenna pointing angles for each TX-RX link, the probability of reception over all RX antenna pointing angles was calculated. For outage discussion purposes, if a signal was recordable above the 5 dB SNR threshold for all 72 RX antenna pointing angles, then the link was not considered an outage for the following analysis. 

Of all 36 TX-RX combinations measured, 20 RX locations resulted in recordable PDPs for all 72 measured RX pointing angles (3 different elevation angles, and all 24 azimuthal angles per elevation angle). This resulted in all random RX locations to effectively receive detectable signal 55.6\% of the time from just a single TX when considering the entire 120$^\circ$ TX sector, over all arbitrary RX antenna pointing directions. The data was then examined to determine the probability of detectable received signal at the RX over all RX antenna pointing angles if being served by two, three, four, or five base stations. For an RX served by exactly two 120$^\circ$ sectored base stations (all combinations of 1 RX and 2 TX base stations), a received signal was detectable over all 72 RX antenna pointing angles, 81.5\% of the time, a nearly 26\% improvement in reception for all beamformed angles compared to service from a single base station. 

\begin{table}
	\centering
	\caption{Probability that an RX experienced signal reception over all antenna pointing angles for one to five serving base stations.} \label{tbl:probs}
	\scalebox{.95}{
		\begin{tabular}{|C{2cm}|C{2.8cm}|C{2.5cm}|}  \hline
			\# of Serving Base Stations & Probability of Reception over all RX Angles & \# of Link Combinations \\ \specialrule{1.5pt}{0pt}{0pt}\centering
			1 & 55.6\%	& 36	\\ \hline
			2 & 81.5\%	& 54	\\ \hline
			3 & 90.5\%	& 42	\\ \hline
			4 & 94.1\%	& 17	\\ \hline
			5 & 100.0\%	& 3		\\ \specialrule{1.5pt}{0pt}{0pt}
		\end{tabular}}
	\end{table}

Next, all possible TX-RX combinations for which an RX is served by three, four, or five base stations were considered for determining service/outage probability. A received signal was detectable over all arbitrary RX antenna pointing angles when served by three base stations with a probability of 90.5\%. This indicates an additional 9\% improvement compared to two serving base stations. For RX's with four serving base stations, only one RX lacked reception over all antenna pointing angles. Additionally, for the 3 combinations where an RX had five serving base stations, a signal was recordable over all RX antenna pointing angles, resulting in no directional outage. Table~\ref{tbl:probs} provides the probabilities of signal reception over all 72 arbitrary RX antenna pointing angles with one to five serving base stations. These probabilities indicate the likelihood of a non-outage over all RX antenna beamformed pointing angles spanning 360$^\circ$ and 45$^\circ$ in the azimuth and elevation planes, respectively, for wide TX sectors of 120$^\circ$. This observation is extremely important at mmWave frequencies and for electrically steerable antenna beams where mobiles will need to be served by multiple APs. The results here indicate that if three base stations serve a single RX to maintain coverage, and if 90.5\% of the time all beamformed angles can maintain a detectable received signal, then in the event of a rapid fade or blockage in one direction, the RX antenna could quickly beam switch to another direction with a high probability of maintaining the link and avoiding loss in coverage. 

\subsection{Nearest Neighbor and Best-N Omnidirectional Path Loss}
\begin{figure}
	\centering
	\includegraphics[width=0.42\textwidth]{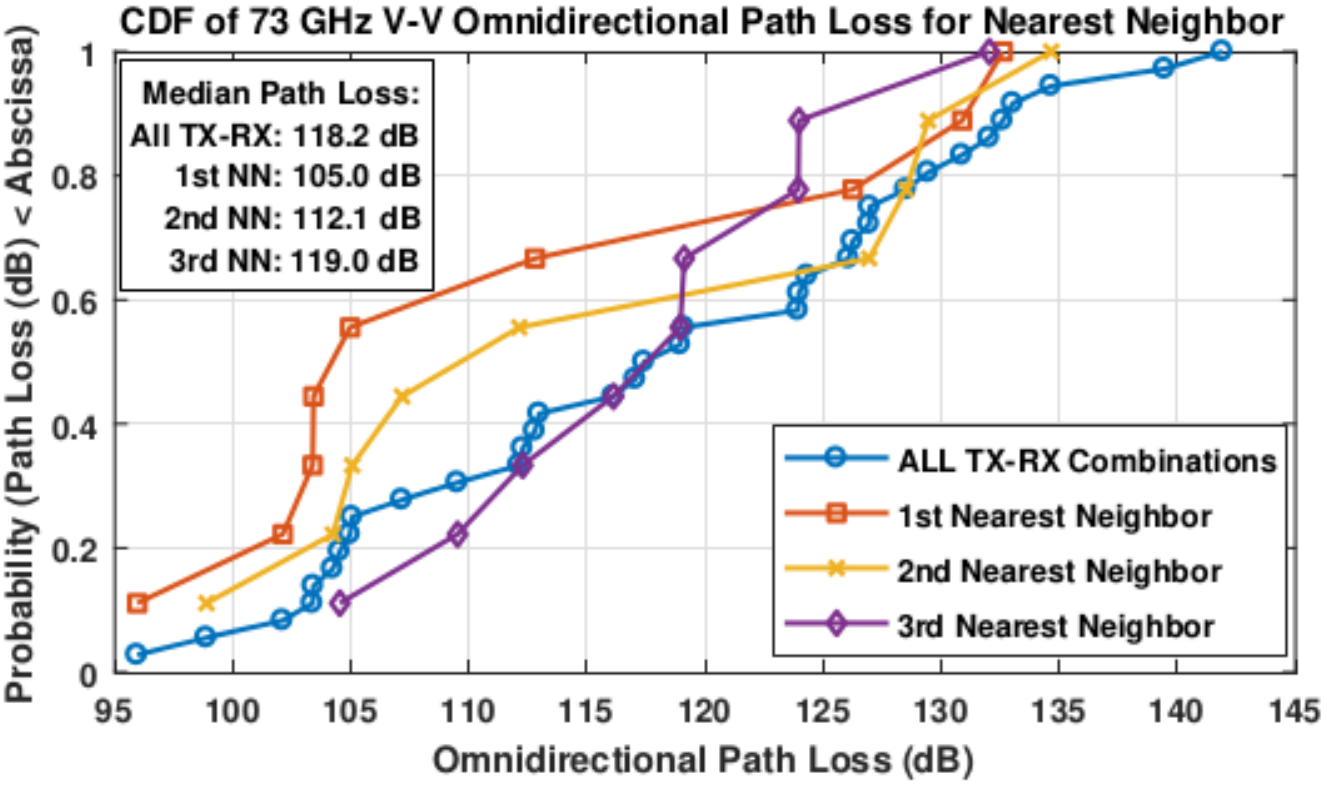}
	\caption{CDF of 73 GHz V-V omnidirectional path loss at an RX for nearest neighbor base stations.}\label{fig:OmniPLNN}
\end{figure}

The CDFs of omnidirectional path loss for nearest neighbors or the closest TX locations in Euclidean distance to each RX location are plotted in Fig.~\ref{fig:OmniPLNN}. Note that there were 9 RX locations with 3 serving base stations and 1 RX location with 1 serving base station. In Fig.~\ref{fig:OmniPLNN} the 1st nearest neighbor (NN) corresponds to path loss observed at an RX for the closest TX, the 2nd NN corresponds to path loss observed at an RX for the second closest TX, and so on. Inspection of the median values shows that the 1st, 2nd, and 3rd NNs are separated by 7 dB each. An interesting note though is that NN does not always result in the lowest path loss, as noticed when comparing the 90\% to 100\% marks on the CDFs.

\begin{figure}
	\centering
	\includegraphics[width=0.42\textwidth]{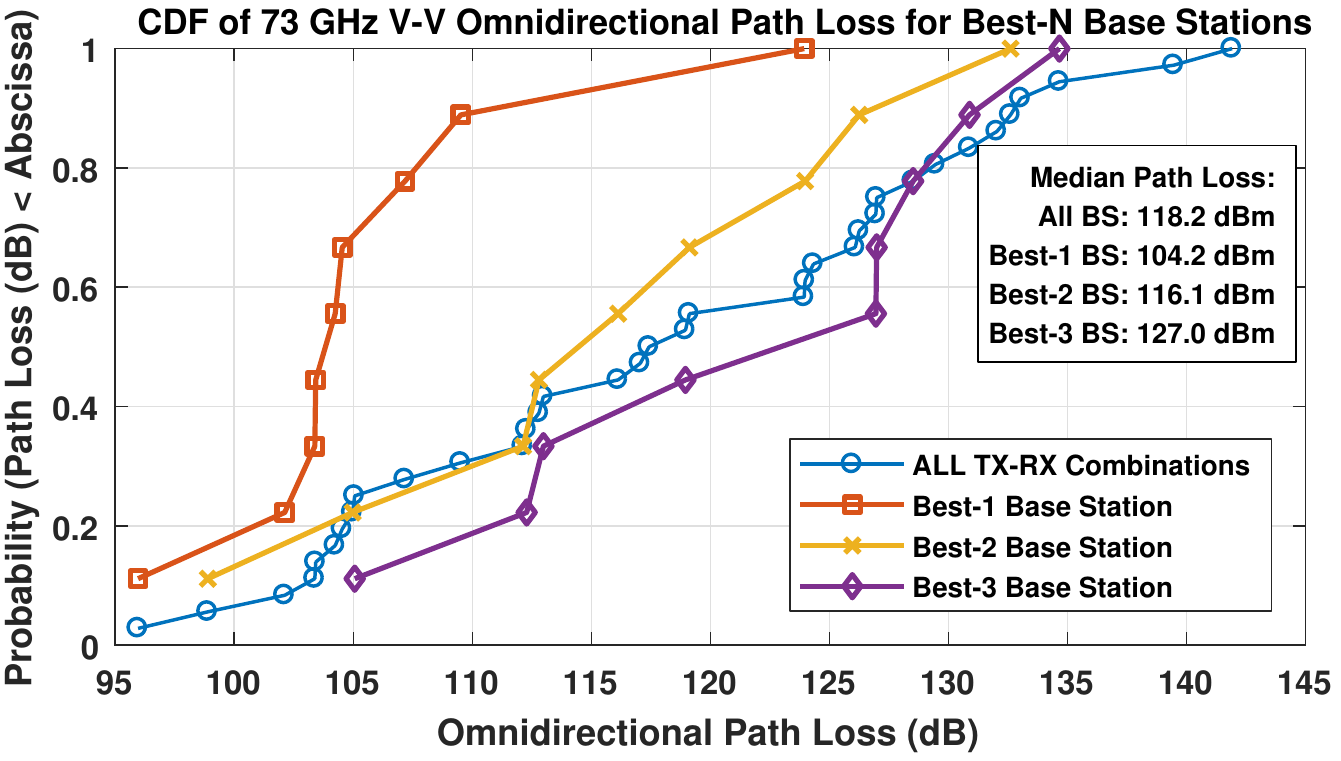}
	\caption{CDF of 73 GHz V-V omnidirectional path loss at an RX for the Best-\emph{N} serving base stations.}\label{fig:OmniPLTopN}
\end{figure}

For the Best-\emph{N} base stations, the omnidirectional path losses for each RX and its serving base stations are sorted from lowest to highest such that \emph{N}=1 considers the lowest path loss for each RX, \emph{N}=2 considers the second lowest path loss for each RX, and so on. The CDFs for the Best-\emph{N} up to \emph{N}=3 are plotted in Fig.~\ref{fig:OmniPLTopN} and a noteworthy observation is the difference in median omnidirectional path loss for the Best-\emph{N} base stations, where the Best-1 and Best-2 median path losses differ by almost 12 dB and the Best-2 and Best-3 base station path losses differ by nearly 11 dB. The CDFs reveal how much interference to expect from the 2nd strongest base station for a given RX, or the coverage one can expect if the strongest base station is blocked or inaccessible. 

\subsection{Coverage Distance and Coverage Area Probabilities}
Using the median nearest neighbor distances in Table~\ref{tbl:NNdistances} along with 100 m and 200 m (predicted mmWave cell radii), a simple signal coverage analysis was performed. For typical system coverage and worst-case analysis, the NLOS and NLOS-best path loss model parameters from Table~\ref{tbl:Dir_PL_VV_CI} were used to determine the probability that a received signal is above the received signal threshold for a mobile at a median nearest neighbor distance or the cell-edge. The system parameters for the measurements were used for analysis (see Table~\ref{tbl:sysSpec}) which resulted in a receiver threshold of:
\begin{equation}\label{eq:PrThresh}
\footnotesize
\begin{split}
P_{r(\thresh.)} &= P_{TX}+G_{TX}+G_{RX}-PL_{\textrm{max}}+10\log_{10}\left(\textrm{BW}[\GHz]\right)\\
&=14.9\,\dBm+27\,\dBi+20\,\dBi-175\,\dB+0\,\dB\\
& = -113.1\,\dBm
\end{split}
\end{equation}
The formula in~\eqref{eq:PrThresh} can be modified for a similar system by scaling the bandwidth relative to 1 GHz, for a different transmit power, for various antenna gains, and/or for a different receiver threshold. If the received power level $x$ at a distance $R$ is represented by a normal random variable with mean $\overline{x}$ and shadow fading standard deviation $\sigma$ (in dB) about the distant dependent mean, and a receiver threshold $x_0$, then the probability that the received signal exceeds the threshold on a cell boundary is~\cite{Jakes74}:
\begin{equation}\label{eq:probDist}
\footnotesize
\begin{split}
P_{x_0}(R)&=P[x> x_0]=\int_{x_0}^\infty p(x)dx=\frac{1}{2}-\frac{1}{2}\textrm{erf}\left(\frac{x_0-\overline{x}}{\sigma \sqrt{2}}\right)
\end{split}
\end{equation}
The CI V-V path loss models in Table~\ref{tbl:Dir_PL_VV_CI} were used to determine the mean path loss $\overline{x}$ at a distance $R$ with the shadow fading standard deviation $\sigma$ (in dB), and receiver threshold $x_0=P_{r(\thresh.)}$, in~\eqref{eq:probDist}. These probabilities are also extended to the probability of outage within a coverage region as described in~\cite{Jakes74}.

\begin{table}
	\centering
	\caption{Best and arbitrary beam pointing outage probabilities for median nearest neighbor base station distances from Table~\ref{tbl:NNdistances}.} \label{tbl:ProbOut}
	\fontsize{8}{8}\selectfont
	\scalebox{0.95}{
		\begin{tabu}{|C{1.4cm}|C{1.0cm}|C{1.0cm}|C{1.0cm}|C{1.0cm}|C{1.0cm}|}  \specialrule{1.1pt}{0pt}{0pt}
			\multicolumn{6}{|c|}{\textbf{$P_{\textrm{outage}}$ at Cell-Edge}} \\ \specialrule{1.1pt}{0pt}{0pt}
			& \multicolumn{3}{c|}{\textbf{Nearest Neighbor}} & \multicolumn{2}{c|}{\textbf{Cell Radius}} \\ \hline
			Scenario & 1st & 2nd & 3rd & 100 m & 200 m\\ \specialrule{1.1pt}{0pt}{0pt}
			NLOS & 2.4\% & 5.5\% & 8.0\% & 12.2\% & 52.0\% \\ \hline
			NLOS-Best & 6.9E-5\% & 2.3E-4\% & 4.1E-4\% & 8.6E-4\% & 2.3E-2\% \\ \specialrule{1.1pt}{0pt}{0pt}
			\multicolumn{6}{|c|}{\textbf{$P_{\textrm{outage}}$ in Region w/ Cell Radius}} \\ \specialrule{1.1pt}{0pt}{0pt}
			& \multicolumn{3}{c|}{\textbf{Nearest Neighbor}} & \multicolumn{2}{c|}{\textbf{Cell Radius}} \\ \hline
			Scenario & 1st & 2nd & 3rd & 100 m & 200 m\\ \specialrule{1.1pt}{0pt}{0pt}
			NLOS & 0.7\% & 1.8\% & 2.8\% & 4.6\% & 27.1\% \\ \hline
			NLOS-Best & 1.8E-5\% & 6.0E-5\% & 1.1E-4\% & 2.4E-4\% & 7.1E-3\% \\ \specialrule{1.1pt}{0pt}{0pt}
	\end{tabu}}
\end{table}

Table~\ref{tbl:ProbOut} provides the NLOS and NLOS-best model probability that a mobile experiences an outage at the median distances from the three nearest base stations as well as the probability of outage within a coverage region having a cell radius of the nearest neighbor base station median distances. The coverage probabilities at a cell-edge in Table~\ref{tbl:ProbOut} show that a mobile may experience only an 8\% outage at a median distance from the 3rd nearest base station in NLOS for any arbitrary antenna pointing direction. A cell with radius 200 m has 52\% outage at the cell-edge, but improves nearly 40\% for a radius of 100 m. Surprisingly, outage at the cell-edge and within a coverage region from the nearest neighbor and max cell radius base stations are nearly zero percent for situations in which the TX and RX antennas can perfectly align their beams to provide maximum received power in NLOS (NLOS-best). This indicates a significant improvement in coverage for mmWave cellular networks that can beamform and beam track under NLOS conditions, even if the 1st and 2nd nearest neighbor base stations are inaccessible. This preliminary study shows macro-diversity and joint transmission or coordinated scheduling CoMP in mmWave small-cells will be advantageous.  

\section{Conclusion}\label{sec:conc}
A new base station diversity and CoMP-style measurement campaign conducted at 73 GHz mmWave was presented. The measurements were conducted in a UMi O.S. scenario in downtown Brooklyn, New York to study the impact of AP diversity and coverage at mmWave bands. In all, 36 total TX-RX location combinations were measured for a number of directional TX and RX antenna pointing angles. LOS and NLOS directional path loss models were derived, indicating that LOS propagation in the UMi O.S. scenario matches perfectly with FSPL with a PLE of 2.0, and minimal shadow fading of 1.9 dB. While the directional NLOS PLE was 4.6 (similar to other UMi campaigns~\cite{Rap15b}), the directional NLOS PLE was 2.9 when considering the best antenna pointing angles between the TX and RX. This is on par with sub-6 GHz NLOS path loss models, as long as the TX and RX can beamform and steer in the optimal directions. Omnidirectional path loss models synthesized from the directional measurements revealed a LOS PLE of 1.9 and a NLOS PLE of 2.8 which are nearly identical to path loss parameters in the literature. 

Probabilities of signal reception over all tested beamformed RX pointing angles for wide sectored TX angles of departure were provided for scenarios with one to five serving base stations. For one serving base station, an RX was able to receive a signal over all antenna pointing angles with a probability of 55.6\%. The probability of signal reception was improved when increasing the number of serving base stations to an RX, resulting in 81.5\% and 90.5\% probability of signal reception across all RX angles when served by two and three base stations, respectively. The nearly 26\% improvement from one to two serving base stations and the 9\% improvement from two to three serving base stations shows that an RX has a high probability of maintaining signal coverage over all angles of arrival in a mmWave small-cell with macro-diversity and multiple APs. Additionally, if service to one arbitrary angle is blocked or experiences a deep signal fade, the results here show that either another arbitrary angle at the RX could be used with high probability, or rapid re-routing handoff between another serving base station could be employed. 

Nearest neighbor and Best-\emph{N} statistics for omnidirectional path loss at the mobile handset were provided for coverage and interference analysis for mmWave macro-diversity and CoMP. An interesting observation was the nearly 12 dB median difference in omnidirectional path loss between the Best-1 and Best-2 serving base stations. Analysis for NLOS showed that a mobile being served by the 3rd nearest neighbor base station will receive a signal above a receiver threshold with a 92\% probability for arbitrary TX and RX antenna pointing angles. Additionally, if the TX and RX antennas can beam steer and optimally align in NLOS, then the probability of outage in a coverage region with a cell radius of 200 m is less than one-hundredth of a percent. Future work will explore spatial lobes and angular spread, temporal delay, and more, from the AP diversity measurements described herein. 
\bibliography{MacCartney_Bibv6}
\bibliographystyle{IEEEtran}
\end{document}